\documentclass[twocolumn,showpacs,preprintnumbers,amsmath,amssymb]{revtex4}

\usepackage{graphicx}
\usepackage{dcolumn}
\usepackage{bm}

\newcommand{\be}{\begin{equation}}
\newcommand{\ee}{\end{equation}}
\newcommand{\bdis}{\begin{displaymath}}
\newcommand{\edis}{\end{displaymath}}
\newcommand{\pd}{\partial}

\begin{document}

\title{False vacuum decay via Coleman - de Luccia instanton in the theory with quasi-exponential potential}

\author{Michal Demetrian}\email{demetrian@fmph.uniba.sk}
\affiliation{Department of Theoretical Physics, Comenius University \\
Mlynsk\' a Dolina F2, 842 48, Bratislava, Slovak Republic }

\date{\today}

\begin{abstract}
The necessary and sufficient condition for the
false vacuum decay in a de Sitter universe via Coleman - de Luccia instanton
is applied on the scalar field with quasi-exponential potential.
\end{abstract}

\maketitle

\section{Introduction}
The theory of false vacuum decay in de Sitter universe via formation of rapidly expanding bubbles
was introduced by Coleman and de Luccia \cite{cdl}. The process has been studied intensively as an
ingredient of old inflation \cite{guth} and of open inflation \cite{{oi1},{oi2}}. This mode of vacuum
decay is described by the Coleman - de Luccia (CdL) instanton - the nontrivial finite-action O(4) solution
to the Euclidean equations for coupled scalar and gravitational fields,
with an effective potential containing a true vacuum (at which it has
the global minimum equal to zero) and a false vacuum (at which it has a
local minimum) separated from the true vacuum by a finite potential
barrier. It is given by two functions $\Phi (\tau)$ and $a (\tau)$, where
$\Phi$ is the scalar field, $a$ is the radius of the three-spheres of homogeneity
determined from the circumference and $\tau$ is the radius of the three-spheres
of homogeneity. Denote the effective potential by $V$. Functions $\Phi$ and $a$ obey equations
\be\label{ee}
\ddot{\Phi} + \frac{3\dot{a}}{a}=V',\
\dot{a}^2=1+\frac{8\pi}{3}\left(\frac{1}{2}\dot{\Phi}^2-V\right)a^2,
\ee
where the overdot
denotes differentiation with respect to $\tau$ and the prime differentiation with respect to $\Phi$.
The boundary conditions ensuring finiteness of the (Euclidean) action
\bdis
S_E= \int{\rm d}^4x\ \sqrt{|g_E|}\left[ \frac{1}{2}g_E(\pd\Phi,\pd\Phi)+V(\Phi)-\frac{R}{16\pi}\right]
\edis
are
\be \label{eebc}
a(0)=\dot{\Phi}(0)=\dot{\Phi}(\tau_f)=0,
\ee
where $\tau_f>0$ is defined by the equation $a(\tau_f)=0$. The existence of a (nontrivial) solution of the
system of equations (\ref{ee},\ref{eebc}) remained unclear for many years. It has been supposed
that the bubble (if it exists) must fit into the sphere, which is the false vacuum  (constant $\Phi$)
solution of the equations (\ref{ee}); and this statement has been translated approximatively into the condition
that the curvature (absolute value of the second derivative) of the potential $V$ in the false vacuum must be bigger
than $4\times$ corresponding Hubble parameter squared.

\section{The necessary and sufficient condition for the existence of CdL instanton}

Being motivated by the arguments of the papers \cite{{hm}} and \cite{js}, the authors of the paper
\cite{vladoaja} have found a necessary condition and a sufficient condition for the existence of the CdL instanton.
Before we formulate these conditions
let us introduce useful notations used in what follows: by
$V_M=V(\Phi_M)$ we denote the local maximum of $V$
(the top of the barrier), and by $H(\Phi)$ the value of the Hubble parameter corresponding to the energy
density equal to $V(\Phi)$, $H(\Phi)=\sqrt{8\pi V(\Phi)/3}$
The conditions read: \\

{\it If the inequality $-V''_M/H_M^2>4$ holds then the CdL instanton exists. } \\

{\it If the CdL instanton exists then the inequality $-V''(\Phi)/H^2(\Phi)>4$ holds somewhere in the barrier. } \\

For a given theory (potential) the conditions leave a
{\it gray zone} in the parameter space of the potential in which
one cannot decide whether the CdL instanton exists or not.

\section{Quasi-exponential potential}

\begin{figure}[h]
\includegraphics[width=5cm,height=4cm]{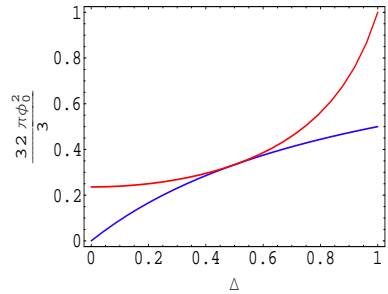}
\caption{The blue (red) curve represents the sufficient (necessary) condition. The area between the curves
is the gray zone for the potential (\ref{pot}).}
\label{f1}
\end{figure}

We are interested in the theory with the potential
\be \label{pot}
V(\Phi)=\left[ V_0+\frac{1}{2}\Phi^2\right]e^{-\Phi/\Phi_0},
\ee
where $V_0$ and $\Phi_0$ are positive parameters. The classical evolution of the scalar field coupled to gravity and
self-interacting by the potential similar to ours has been discussed in many works; as an example and source of
further information we refer to the paper \cite{barrow}. In order to apply our conditions to the theory
(\ref{pot}) we have to do nothing but investigate elementary properties of the function
$V(\Phi)$. The barrier exists only if $V_0<1/2\Phi_0^2$. If we introduce the new parameter
\bdis
\Delta=\sqrt{1-2\frac{V_0}{\Phi_0^2}},\quad \Delta\in [0,1],
\edis
the formula (\ref{pot}) becomes
\bdis
V(\Phi)=\frac{1}{2}\left[ \Phi^2+\Phi_0^2(1-\Delta^2)\right]e^{-\Phi/\Phi_0} .
\edis
The local extremes of the function $V$ are reached at the points
\bdis
\Phi_{m,M}=\Phi_0(1\mp\Delta)\ \mbox{and}\ V_M=\Phi_0^2(1+\Delta)e^{-1-\Delta}.
\edis
The sufficient condition holds only if
\be \label{scexp}
\frac{32\pi}{3}\Phi_0^2<\frac{\Delta}{\Delta+1} .
\ee
In words, the sufficient condition holds if the barrier is sufficiently thin. The parameter $\Delta$
measures the thickness of the barrier because,
if $\Phi_+$ and $\Phi_-$ are
the inflexion points of $V$ to the right and left of $\Phi_M$, we have
$\Phi_+-\Phi_-=2\Phi_0\sqrt{1+\Delta^2}$.
The necessary condition is fulfilled only if the curves $32\pi V(\Phi)/3$ and $-V''(\Phi)$ have at least
one common point in the interval $[\Phi_-,\Phi_+]$ (The curves can have two intersection points in that
interval).
After a little bit more complicated algebra one gets that this is true only if
\be \label{ncexp}
\frac{32\pi}{3}\Phi_0^2<
\frac{\sqrt{5-4\Delta^2}+\Delta^2-2}{1-\Delta^2}.
\ee
The inequalities (\ref{scexp},\ref{ncexp}) are depicted in the figure \ref{f1}.

\acknowledgments{The author thanks Vlado Balek for useful discussions. This paper was supported by the
grant VEGA 1/0250/03.}

\end{document}